# STACKED PATCH ANTENNA WITH CROSS SLOT ELECTRONIC BAND GAP STRUCTURE

N.S. Raghava[1], Asok De[2], Nitish Kataria[3], Sarthak Chatterjee[4]

[1, 2] Dept. Of Electronics and Communication Engg., Delhi Technological University, Delhi, India

[3, 4] Dept. Of Electronics and Communication Engg., GPMCE, IP University, Delhi, India

## ABSTRACT

**A cross slotted electronic band gap (EBG) with stacked rectangular patches shorted with a shorting pin is proposed in this paper. The study is being done on how the various parameters are varied by changing the probe feed location. The design is constructed by using stacking of patches, shorting pin and cross slotted EBG to form an optimized antenna design with antenna efficiency of approximately 99.06%. The radiation patterns are given at 2.586 GHz which can be used for wireless communications.**

## 1. INTRODUCTION

As the wireless mobile communication technologies is growing very rapidly, microstrip antenna have various applications because of their small size, low weight, low cost, low planar configuration and can be easily integrated with other microwave circuits. They are very compact and wideband antennas for the transmission of video, voice and data information. The various applications for microstrip antennas are wireless communication systems, satellite communication systems, cellular phones, pagers, radars, etc.

The 2.6 GHz band is generally considered to cover the frequency range between 2500-2690MHz, although there are some minor national variations. The band provides an opportunity to meet rapidly rising demand for capacity to deliver mobile broadband services on widespread, common basis across the world. The applications of the frequencies in this band, which vary by country and region, includes satellite services, fixed mobile and broadcast as well as terrestrial and video broadcasting in certain countries. In this paper, the frequency of 2.586GHz is used which can be implemented in GPRS, WLAN, mobile broadcasting and road vehicle communications (Raghava and Asok De, 2006).

Small antenna design is always compromised between size, bandwidth and efficiency. The bandwidth can be increased with the use of stacked patch structure in a microstrip antenna (MSA) (Ollikainen, Fischer and Vainikainen, 1999) design whereas the size of the antenna can be decreased by the use of shorting pin. The stacked structure can be shorted with the use of shorting pin and the overall size of the antenna is reduced drastically (Sanad, 1994). The radiation characteristics can be further improved by an electronic band gap structure on one side of the substrate. This reduces the surface waves induced in the antenna and also increases the radiation pattern (Gonzalo, MAagt, Sorolla, 1999). The use of the reflector plane is fed on the finite size ground plane at the rear end of the antenna to reduce the level of back radiations (Raghava and Asok De, 2009).

In this paper, a stacked rectangular microstrip antenna using a shorting post backed with Cross Slot Electronic Band Gap (CSEBG) structure is studied. The effect on antenna efficiency and gain is also compared by changing the probe feed location while the shorting post remained fixed.

## 2. ANTENNA DESIGN

A stacked rectangular microstrip antenna using a shorting post backed with Cross Slot EBG structure (CSEBG) is designed. The top view of typical EBG stacked rectangular microstrip antenna with a shorting post is shown in Fig. 1. The antenna consists of two dielectric layers with stacked rectangular patches of dimensions (16 x 16) mm. The thickness of the upper dielectric (i) having dielectric constant of air is 2.6mm and that of the lower one (ii) having dielectric constant 2.6 is 1.6mm. The square patches used are mounted over the substrate creating a stacked structure. The stacking of patches increases the antenna efficiency and gain of the microstrip antenna. The 3-D view of CSEBG MSA is given in Fig. 2.

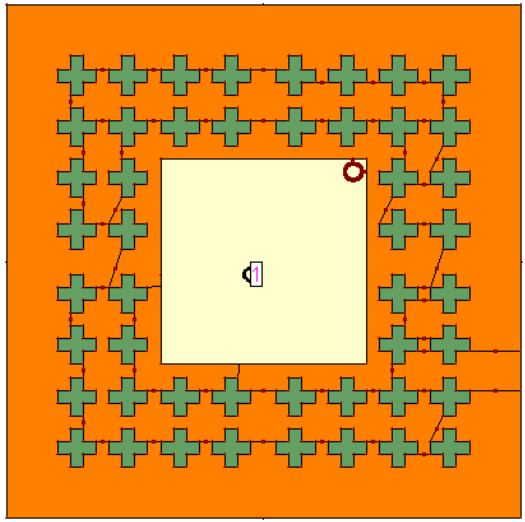

**Fig. 1** *Top View of CSEBG MSA*

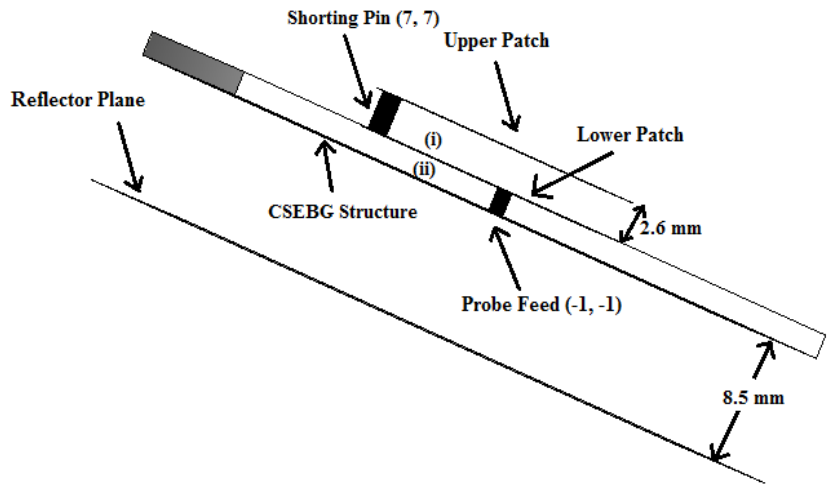

**Fig. 2** *3-D view of CSEBG MSA*

The EBG structure is constructed by etching cross shaped slots on a square patch of size (40 x 40) mm. The cross shaped slots of dimensions (3 x 1) mm which are spaced 1 mm apart forms an 8 x 8 matrix. The distance from the edge is 4mm and there are no slots beneath the patch area. The EBG structure is used for better radiation characteristics.

This antenna is excited by a coaxial probe feed of diameter 1mm given at location (-1, -1). The two rectangular patches are shorted by a shorting post of diameter 1.3mm at location (7, 7). The use of shorting post decreases the size of the antenna (M.Sanad, 1994). A reflector plane is placed at a distance of 8.5mm behind the ground plane as given in Fig. 2. The reflector plane is used to reduce the back lobes generated by the antenna. In this paper, the various parameters like antenna efficiency, gain and directivity are improved by using stacked rectangular patches and shorting them with shorting post backed with cross slotted EBG structure (CSEBG). These parameters are also compared at different probe feed locations. The proposed CSEBG MSA at probe feed location (-1, -1) with antenna efficiency of 99.06 % and gain 4.75 dB is able to work at 2.586 GHz frequency and can be used for various wireless communication application.

## 4. RESULTS AND DISCUSSIONS

The CSEBG MSA is simulated using IE3D simulator which is based on the method of moments (MoM) (Zeland, ver. 14.0, 2007). We have done a comparison of antennas various parameters at different locations of probe feed. The table 1 shows that the antenna efficiency and gain for CSEBG MSA at location (-1,-1) are approximately 99.06% and 4.75 dB which are highest with respect to others. The antenna parameters like antenna efficiency and gain show strange behaviour when the probe feed location is near shorting post as the parameters decrease drastically as we approach it. The 3-D radiation pattern for CSEBG MSA is shown in Fig. 3. The Fig. 4 shows the graph between frequency and return loss (S11) and it can be seen that highest negative return loss -19.96 is at frequency 2.586 GHz. The resonant frequency of CSEBG MSA is 2.586 GHz which has various applications in wireless and road vehicle communications.

**Table. 1** *Comparison at different probe feed locations*

| S.No. | Location Of Probe Feed | Frequency (GHz) | Radiation Efficiency (%) | Antenna Efficiency (%) | Gain (dB) | Directivity (dB) | Return Loss (S11) |
|---|---|---|---|---|---|---|---|
| 1 | (-7,-7) | 2.616 | 100 | 97.1155 | 4.5495 | 4.67658 | -15.29 |
| 2 | (-6,-6) | 2.606 | 100 | 96.9724 | 4.5354 | 4.66893 | -15.22 |
| 3 | (-5,-5) | 2.606 | 100 | 97.337 | 4.5748 | 4.69197 | -15.76 |
| 4 | (-4,-4) | 2.606 | 100 | 97.7917 | 4.6228 | 4.71975 | -16.53 |
| 5 | (-3,-3) | 2.6 | 100 | 98.3209 | 4.6612 | 4.73471 | -17.75 |
| 6 | (-2,-2) | 2.586 | 100 | 98.8704 | 4.7254 | 4.77476 | -19.64 |
| **7** | **(-1,-1)** | **2.586** | **100** | **99.0584** | **4.7488** | **4.78993** | **-19.96** |
| 8. | (1,1) | 2.556 | 100 | 98.0513 | 4.7814 | 4.86684 | -16.81 |
| 9 | (2,2) | 2.606 | 100 | 87.5899 | 4.2951 | 4.87054 | -9.056 |
| 10 | (3,3) | 2.6 | 100 | 73.3518 | 3.5751 | 4.92096 | -5.743 |
| 11 | (4,4) | 2.596 | 100 | 55.5322 | 2.4456 | 5.00016 | -3.521 |
| 12 | (5,5) | 2.576 | 100 | 11.6286 | -4.204 | 5.14061 | -0.5345 |

## 5. CONCLUSIONS

The cross slotted EBG with stacked rectangular microstrip antenna (CSEBG MSA) is proposed in this paper. The antenna efficiency and gain are 99.06 % and 4.75 dB respectively. The probe feed locations are changed along the diagonal. The different values of antenna parameters are compared at different probe feed locations. The highest value of antenna efficiency and gain is at location (-1, 1) but as we approach towards shorting post their values decrease drastically. The use of stacked structure and shorting them with shorting pin also increases the gain and efficiency of the antenna. For the calculation of radiation efficiency no losses are taken into account, whereas in the case of antenna efficiency losses due to mismatch, ohmic and dielectric etc. are taken into account. With good antenna efficiency and gain of the proposed antenna at frequency 2.586 GHz, it can be used for GPRS, WLAN and road vehicle communications.

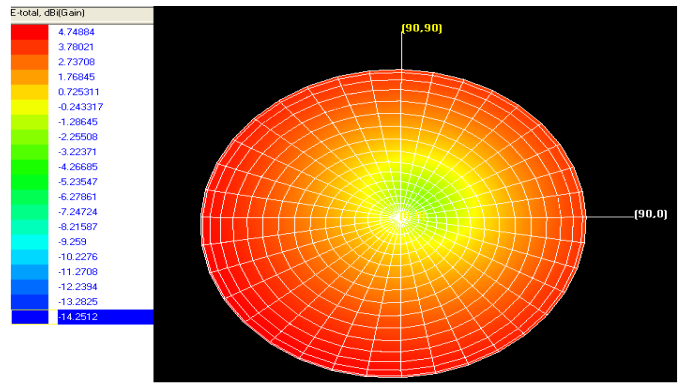

**Fig. 3** *3-D Radiation Pattern for CSEBG MSA*

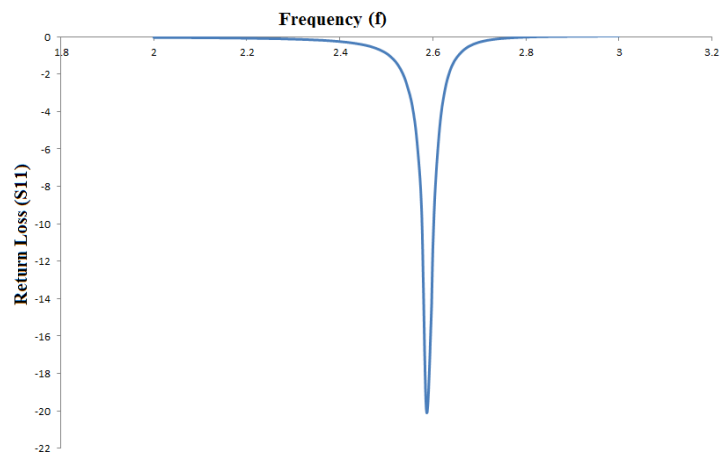

**Fig. 4** *Plot between return loss (S11) and frequency of CSEBG MSA*